\documentclass[11pt,a4paper]{article}

\def\msusy{\ensuremath{M_\text{SUSY}}}
\def\mgut{\ensuremath{M_\text{GUT}}}

\usepackage{jheppub}
\usepackage{subfigure}
\usepackage{epstopdf}
\usepackage{feynmp}
\author[a]{B.~C.~Allanach}
\affiliation[a]{Department of Applied Mathematics and Theoretical Physics, Centre for Mathematical Sciences, University of Cambridge, Wilberforce Road, Cambridge CB3 0WA, United Kingdom}
\emailAdd{B.C.Allanach@damtp.cam.ac.uk}
\author[a,b]{Damien P.~George}
\emailAdd{dpg39@cam.ac.uk}
\author[b]{Ben Gripaios}
\affiliation[b]{Cavendish Laboratory, University of Cambridge, JJ Thomson
  Avenue, Cambridge CB3 0HE, United Kingdom}
\emailAdd{gripaios@hep.phy.cam.ac.uk}

\title{The dark side of the $\mu$: on multiple solutions to renormalisation
  group equations, and why the CMSSM is not necessarily being ruled out}

\abstract{When solving renormalisation group equations in a quantum field
  theory, one often 
  specifies the boundary conditions at multiple renormalisation
  scales, such as the weak and grand-unified scales in a theory beyond the
  standard model.  A point in the parameter space of such a model is usually
  specified by 
  the values of couplings at these boundaries of the renormalisation group flow,
  but there is no theorem guaranteeing that such a point has a unique solution
  to 
  the associated differential equations, and so there may exist multiple,
  phenomenologically distinct solutions, all corresponding to the same
  point in parameter space.  
  We show that this is indeed the case in the constrained minimal supersymmetric
  standard model (CMSSM), and we exhibit such solutions, which cannot be
  obtained using out-of-the-box computer programs in the public domain. 
  Some of the multiple solutions we exhibit have CP-even lightest Higgs mass
  predictions between 124 and 126 GeV.  
  Without an exhaustive
  11-dimensional MSSM parameter scan per CMSSM parameter point to capture
  all of the multiple solutions, CMSSM phenomenological analyses are
  incomplete.} 

\begin{document}
\maketitle

%%%%%%%%%%%%%%%%%%%%%%%%%%%%%%%%%%%%%%%%%%%%%%%%%%%%%%%%%%%%%
\section{Introduction}
%%%%%%%%%%%%%%%%%%%%%%%%%%%%%%%%%%%%%%%%%%%%%%%%%%%%%%%%%%%%%

It is a truth universally acknowledged, that renormalisation group (RG)
flows are unique, once a boundary condition for each coupling involved
in the flow has been specified.
%in the sense that if we specify all couplings at
%some scale and then flow to another scale, the resulting trajectory
%will be unique.

Like many universally acknowledged truths, this one is not
necessarily acknowledged in the Universe inhabited by mathematicians. Indeed, the closest one can get in
terms of a theorem (called Cauchy--Lipschitz by francophones, but due to
Picard--Lindel\"{o}f \cite{Lindelhof}) concerns the uniqueness
(and existence) of the solution to the initial value problem of a sufficiently
well-behaved system of differential equations in a neighbourhood
of the starting point. There are many situations in physics where these
conditions are not 
satisfied and so the issue of non-uniqueness (as well as non-existence)
rears its ugly head. 

One physical situation where non-uniqueness is
manifest arises in
Sturm-Liouville
problems, namely linear, second-order, ordinary differential equations (ODEs)
with homogeneous boundary conditions specified on either side of an
interval. These may be thought of as a pair of first-order ODEs and, as every undergraduate knows, there exists either 1 (namely 0) or
infinitely many solutions (namely an eigenfunction multiplied by an
arbitrary constant) depending on whether the ODEs
correspond to an eigenvalue or not.

A different example, involving RG flows, is pertinent right now at the Large
Hadron Collider (LHC). Disciples of high-scale supersymmetric models (rapidly
becoming an endangered species) wish to know whether
their models are ruled out or not. These models typically
impose a large degree of unification of the parameters of {\em e.g.}
the Minimal Supersymmetric Standard Model (MSSM) at a high energy
scale, be they gauge couplings, soft supersymmetry (SUSY) breaking masses, or
whatever.  Such constraints play the r\^{o}le of boundary conditions
at one end of the RG flow.  The MSSM is then further constrained at the weak
scale where various Standard Model measurements, such as the mass of the Z boson,
also play the r\^{o}le of boundary conditions.  The RG equations (RGEs) are
non-linear ODEs, for which a solution satisfying both sets of boundary
conditions 
is to be found.

For concreteness, let us imagine an
$n$-parameter flow (with $n$ of $O(10^2)$ in the MSSM), in which $k$
measurements are performed at low energy, and there are $l$ unification
conditions at high energy. One may then attempt to answer the question of whether the theory is
ruled out or not in the following way:
choose values of $n-k-l$ of the unified parameters at high energy, and
call those values a `point in the parameter space of the model'. Given that $n$ boundary conditions have now been specified, and invoking the
aforementioned universally-acknowledged truth that the flow is unique, one may find
`the' flow that satisfies the given boundary conditions, by means of a numerical iterative
algorithm.\footnote{In a nutshell, such an algorithm works by choosing initial
guesses for the {\em a priori}\/ unknown parameters at, say, low energy
in order 
to create an artificial initial value problem, which can be solved to
find high energy values of all $n$ parameters. Those which are known
at high energy are discarded in favour of their known values, and one
flows back again. The process is then repeated indefinitely, in the
hope of converging on a solution.}
The resulting flow then predicts the
values of all particle masses and couplings of the theory, including
those which have yet to be measured, but are subject to limits
({\em e.g.}\/ limits on superpartner masses from LEP and LHC). If the limits are
violated, the point in parameter space is ruled out. Finally, one can scan
over points in parameter space {\em ad nauseam}. 

The problem with this approach is that no theorem guarantees that a
solution found in this way is unique. (Indeed, the similarity with a
Sturm-Liouville problem suggests the contrary.) Thus, there may be more than one trajectory satisfying the boundary conditions, each of which
reproduces the unification conditions and the measured Standard Model
parameters, but each of which may have completely different values of
the as-yet unmeasured low-energy parameters. Some of these may be
ruled out and some may not.

This raises the spectre (horrifying or enchanting depending on one's
spiritual taste) that points in the parameter space of, {\em
  e.g.}, the 
CMSSM that have previously been ruled out, are in fact not ruled out at
all, because there are multiple RG flows that correspond to the same
point in CMSSM input parameter space, with one or more flows still allowed
and yet not found by existing algorithms.

In what follows, we shall find multiple RG flows corresponding to the same
CMSSM input parameter point.\footnote{Recently, Ref.~\cite{Liu:2013ula}
  investigated different `branches' of 
electroweak symmetry breaking in the CMSSM (and other models) given current
constraints. However, these are not different {\em multiple solution}\/ branches: 
for a given CMSSM point, only one solution was found. 
} Though
our numerical calculations hint that such points may be uncommon,
it will become clear that we have no way of exhaustively categorising
them.
This, unfortunately, makes it rather difficult to say whether some regions of
some high scale SUSY breaking scheme, 
{\em e.g.}, the CMSSM, are ruled out or not.

Before doing all this, we first try to convince the reader, in
\S~\ref{sec:toy}, of the existence of multiple solutions for a rather more
simple RG flow, 
namely the flow in the neighbourhood of the
Berezinski-Kosterlitz-Thouless (BKT) phase transition. Then, in
\S~\ref{sec:cmssm}, we exhibit and investigate the phenomenon for the CMSSM. 
It is not our purpose here to perform any detailed
phenomenology. Rather, we content ourselves with
pointing out the existence and prevalence of multiple solutions in the CMSSM
parameter space and investigating a few of their properties. 
We discuss the implications and context of our results in \S~\ref{sec:disc}. 

%%%%%%%%%%%%%%%%%%%%%%%%%%%%%%%%%%%%%%%%%%%%%%%%%%%%%%%%%%%%%
\section{Toy example: The BKT phase transition \label{sec:toy}}
%%%%%%%%%%%%%%%%%%%%%%%%%%%%%%%%%%%%%%%%%%%%%%%%%%%%%%%%%%%%%

To find a tractable example of the phenomenon of multiple solutions,
we consider a system of RGEs with just two couplings, which is the
minimal case in which the notion arises of imposing boundary conditions at different scales.

A suitable example is the RG flow corresponding to the
BKT phase transition \cite{Berezinsky:1970fr,Kosterlitz:1973xp}. This phase
transition was first studied in the context of the 2-d XY model in
condensed matter physics and can be used to describe superfluid films
and arrays of Josephson junctions.
It also appears in particle physics in the deconfinement phase
transition in compact $U(1)$ gauge theory in $2$ space dimensions, where
Polyakov showed long ago \cite{Polyakov:1976fu} that vortex
configurations of the gauge field lead to
confinement at zero temperature. The finite-temperature deconfinement
phase transition was shown to be of BKT type by embedding in the
Georgi-Glashow model \cite{Svetitsky:1982gs,Agasian:1997wv} and
also by a variational analysis \cite{Gripaios:2002xb}.

The BKT phase transition is rather special in that it exhibits a line of fixed points ending in a
critical point; in the
neighbourhood of this point, the RGEs may be written in the
form
\begin{align} \label{eq:bkt}
\frac{dx}{dt} &= y^2, \nonumber \\
\frac{dy}{dt} &= xy.
\end{align}

These equations may be solved easily enough by a mathematical beginner, but physicists who
know more than is good for them may find
it amusing to convert the problem to a mechanical one, in the
following way. The RGEs do not form a Hamiltonian system (since
$\partial_x y^2 \neq -\partial_y xy$),
but may be turned into one by the non-canonical transformation
$y = e^z$. Then $\dot{z} =x$ and $\dot{x} = e^{2z}$, such that we may think
of a particle of unit mass and momentum $x$ moving in the potential
$-\frac{1}{2}e^{2z}$. The Lagrangian is $t$-independent and the
conserved energy is, {\em \'{a} la}\/ N\"{o}ther, $E = \frac{1}{2} (x^2 - e^{2z})
= \frac{1}{2} (x^2 - y^2)$. In other words, the trajectories are hyperbolae in
the $(x,y)$ plane. Moreover,
$t$-translation invariance implies that the conditions for the existence of solutions to the boundary
value problem can depend only on the
difference, $t_1 - t_0$, between the initial and final times.

There is, in addition, a symmetry of the equations of motion that is not
a symmetry of the Lagrangian, and therefore does not imply a conserved
charge. This symmetry is the rescaling $(x,z,t)
\rightarrow (\alpha x, z+\alpha, \alpha^{-1} t)$. This symmetry
implies that we can, without loss of generality, always rescale a finite time interval to
be unity, $t_1 - t_0 = 1$.

\begin{figure}\begin{center}
\includegraphics[width=5in]{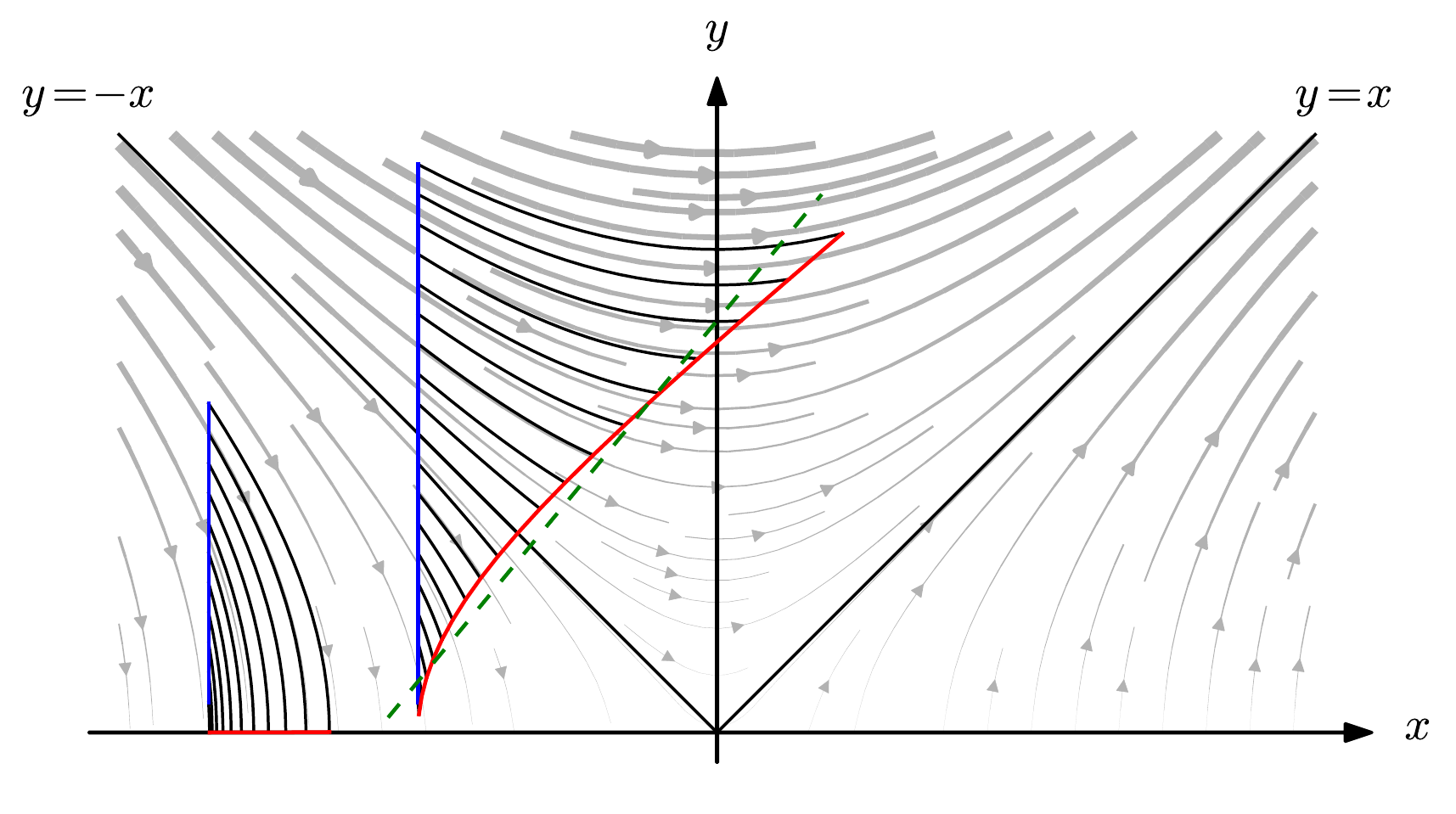}
\caption{Phase diagram and RG trajectories for the BKT system.
  The equations enjoy a scale invariance, so the figure has the same
  character at all scales.  The phase diagram is indicated by the
  background flow-lines (in grey).  We show (in black) two distinct sets
  of flows, with each beginning at a fixed value of $x$ on a blue line,
  and ending after a fixed interval of time, $\Delta t$, on a red line.
  For the flows ending on the curved red line, $\Delta t$ is finite, and
  multiple solutions are indicated by the intersection of this red line
  with the straight, dashed green line.
  For the flows ending on the horizontal red line, $\Delta t$ is infinite.
  \label{fig:bkt}}
\end{center}
\end{figure}

The form of the potential, $-\frac{1}{2}e^{2z}$, indicates that the
momentum, $x$,
must increase monotonically with $t$. There are thus 3 types of
trajectory, as follows. Trajectories with $x(t_0)<0$ and
positive energy, $x^2 - y^2 >0$, proceed inevitably to $z \rightarrow
-\infty$, {\em i.e.} $y=0$, in time $-\int^{-\infty}
\frac{dz}{\sqrt{2E + e^{2z}}} \rightarrow \infty$. Trajectories with $x(t_0)<0$ and $x^2 - y^2 < 0$ eventually turn
around and proceed to $z \rightarrow
+\infty$ in time $\int^{\infty}
\frac{dz}{\sqrt{2E + e^{2z}}} < \infty$, as do trajectories with
$x(t_0)\geq 0$.
The trajectories are thus as shown in grey in
Fig.~\ref{fig:bkt}. 
%%%%%%%%%%%%%%%%%%%%%%%%%%%%%%%%%%
%%%%%%%%%%%%%%%%%%%%%%%%%%%%%%%%%%

It is apparent from the above discussion that the initial value problem is well
defined for any finite starting point $(x(t_0), y(t_0))$ and finite
time interval, but we wish
to consider instead the mixed boundary value problem with, say,
$x(t_0)=-a$
and $y(t_1)=b$. It is immediately evident that the existence of a line of
fixed points permits infinitely many solutions to the boundary value
problem with $b=0$
on an infinite interval: all trajectories with $x(t_0)=-a, y(t_0) < a$
will arrive at the fixed line $y(t_1 \rightarrow \infty) = 0$.  This situation
is depicted in Fig.~\ref{fig:bkt} by the mapping of the smaller blue line to
the horizontal red line.

We are more interested in multiple solutions on a finite time
interval. One can show (by straightforward, but tedious, consideration
of the explicit solutions to Eq.~\eqref{eq:bkt}) that these never arise
for boundary conditions of the specific form $x(t_0)=-a$
and $y(t_1)=b$.
To find multiple solutions on a finite time interval, we need to
consider more general boundary conditions. Consider, for example,
boundary conditions in which we fix $x$ at $t_0$, and fix some
(inhomogeneous) linear combination of $x$ and $y$ at time $t_1$,
say $\alpha x(t_1) + \beta y(t_1) =c$. It is then easy to see that
multiple solutions must arise for particular choices of $\alpha$,
$\beta$ and $c$. Indeed, consider all flows that satisfy $x(t_0)=-a$,
for some other value of $a$, as
per the longer blue line in Fig~\ref{fig:bkt}. If the flow were
linear, the straight line $x(t_0)=-a$ would be mapped into another straight
line at $t=t_1$, which would intersect once with the linear combination appearing in
the boundary condition. (In special cases such as homogeneous boundary
conditions, $a=c=0$, there will be 1 or infinitely many intersections,
as for a Sturm-Liouville problem.) 
But since the flow is
non-linear, the line $x(t_0)=-a$ is mapped to a
curve 
(shown in red), which will intersect multiple times with a suitably chosen
straight line (shown in green)
corresponding to the boundary condition at $t=t_1$. Since, we have
effectively converted the boundary value problem into an initial value
problem for each point on the blue line, for which the RG flow is known to be unique, the number of
solutions to the original boundary value problem is simply given by
the number of intersections of the red curve with the green line.

Notice how, in this way, we convert the problem of finding multiple
solutions of a system of differential equations (the RGEs) into a problem of
finding multiple solutions of a system of algebraic equations (namely
the intersections of the line and curve in
Fig.\ref{fig:bkt}). This makes it relatively straightforward to
establish the existence of multiple solutions for a more general
system (though we still need to
be able to solve the non-linear RGEs of the modified problem in order
to do so): we relax one final, say, boundary condition, find the
resulting flow for many initial values of a
coupling that is not fixed initially, and finally reject flows that
do not satisfy the final boundary condition of the original
problem. If multiple flows remain, then the original problem has
multiple solutions.

It is a much harder problem, in general, to count the total
number of solutions
for a given system of RGEs and boundary conditions. It can be done, in
principle,
by relaxing all of the final boundary conditions and instead scanning
over all possible initial values of parameters that are not
fixed by the initial boundary conditions of the original problem. In
this way, we convert the original problem to an initial value problem,
for which uniqueness of the flow is guaranteed (assuming the equations
are sufficiently well-behaved) by a theorem.

While this is certainly easy enough for a 2-parameter 
flow as above, it is out of the
question for the CMSSM in which, as we will see, we would have to scan over 11 GUT-scale MSSM parameters for a single point in CMSSM parameter space. Thus we cannot be sure of finding all solutions,
and therefore it is essentially impossible for
us to conclusively declare that a generic point in the CMSSM parameter space is ruled
out, in the absence of a theorem on the number of solutions.

The CMSSM is, moreover, further complicated by three features. Firstly, there are really three `endpoints' to the flow, namely the GUT, SUSY, and weak
scales. Secondly, some of the boundary conditions simultaneously specify the
locations of the endpoints themselves, as well as the values that the
Lagrangian parameters take at those endpoints. Thirdly, the boundary
conditions are themselves non-linear relations among the Lagrangian
parameters, meaning that multiple solutions could arise even if the
RGEs were linear.\footnote{In the BKT case, where the RGEs
  are non-linear, but the boundary conditions (BCs) are linear, the existence of multiple
  solutions can be attributed unambiguously to the RGEs. Similarly, if
 the RGEs are linear and the BCs are non-linear, multiple solutions
 can be said to arise from the BCs. But in cases like the CMSSM, where
 both RGEs and BCs are non-linear, there is no sense in which multiple
 solutions can be attributed to one or the other.} Nevertheless, we
can identify multiple solutions easily enough in the following way.
Given an algorithm that finds one solution, relax one of the boundary
conditions (as we did for the BKT example) and scan over one of the couplings appearing in it, finding one solution for each point. If
more than one of these points satisfies all of the original BCs, then
the original problem has multiple solutions.

%%%%%%%%%%%%%%%%%%%%%%%%%%%%%%%%%%%%%%%%%%%%%%%%%%%%%%%%%%%%%
\section{Multiple solutions in the CMSSM \label{sec:cmssm}}
%%%%%%%%%%%%%%%%%%%%%%%%%%%%%%%%%%%%%%%%%%%%%%%%%%%%%%%%%%%%%

The CMSSM~\cite{Fayet:1,Fayet:2,Fayet:3,Fayet:4,Georgi:1}
 is a softly broken model of supersymmetric field theory, 
with boundary conditions at three scales and is therefore subject to the
possibility of several discrete solutions, even though the number of boundary
conditions plus the number of input parameters is equal to the number of free
parameters. It remains the
phenomenologically most studied example of assumptions about supersymmetry
breaking terms in the MSSM, and is still of high interest.
In this section, we will
find some of its multiple solutions, map out the regions of parameter
space in which we can find multiple
solutions, and
illustrate their properties. However, as we have already discussed, we
can by no means
guarantee that we have found {\em all}\/ of its solutions. 

The CMSSM has the following parameters: $\tan \beta$,
the ratio of the two Higgs vacuum expectation values; sign$(\mu)$, the 
sign of a parameter in the MSSM superpotential; 
$m_0$, which is equal to the SUSY
breaking family and flavour universal scalar mass terms in the Lagrangian;
$M_{1/2}$, which is equal to the gaugino mass SUSY breaking mass
parameters; and
 $A_0$, which sets the SUSY breaking trilinear scalar couplings (these are
equal to the analogous Yukawa matrix multiplied by $A_0$). 
The soft SUSY breaking terms are nearly all set at
the GUT scale, which is defined to be the scale where the gauge couplings are
unified, typically $\sim 2-3 \times 10^{16}$ GeV. 
Usually, one expects to obtain
two solutions: one for each sign of $\mu$, although occasionally one runs up
against physical boundaries (such as an unstable electroweak minimum), and one
or both of these solutions is unphysical. 

In fact, multiple solutions in the mSUGRA
model~\cite{Cremmer:1978iv,Cremmer:1978hn,Barbieri:1982eh} have already been
found by Drees and 
Nojiri~\cite{Drees:1991ab}. The mSUGRA model is equivalent to the CMSSM with
one additional constraint on a Higgs potential parameter:
$m_3^2(\mgut)=\mu(\mgut) ( A_0 - m_0 )$. Drees and Nojiri used
this additional constraint to predict $\tan \beta$ 
from the minimisation of the weak scale Higgs potential. The resulting equation
was analytically 
shown to be a cubic in $\tan \beta$, which may have up to three physically
distinct solutions. 

\subsection{CMSSM boundary conditions}
\begin{figure}\begin{center}
\includegraphics[width=5in]{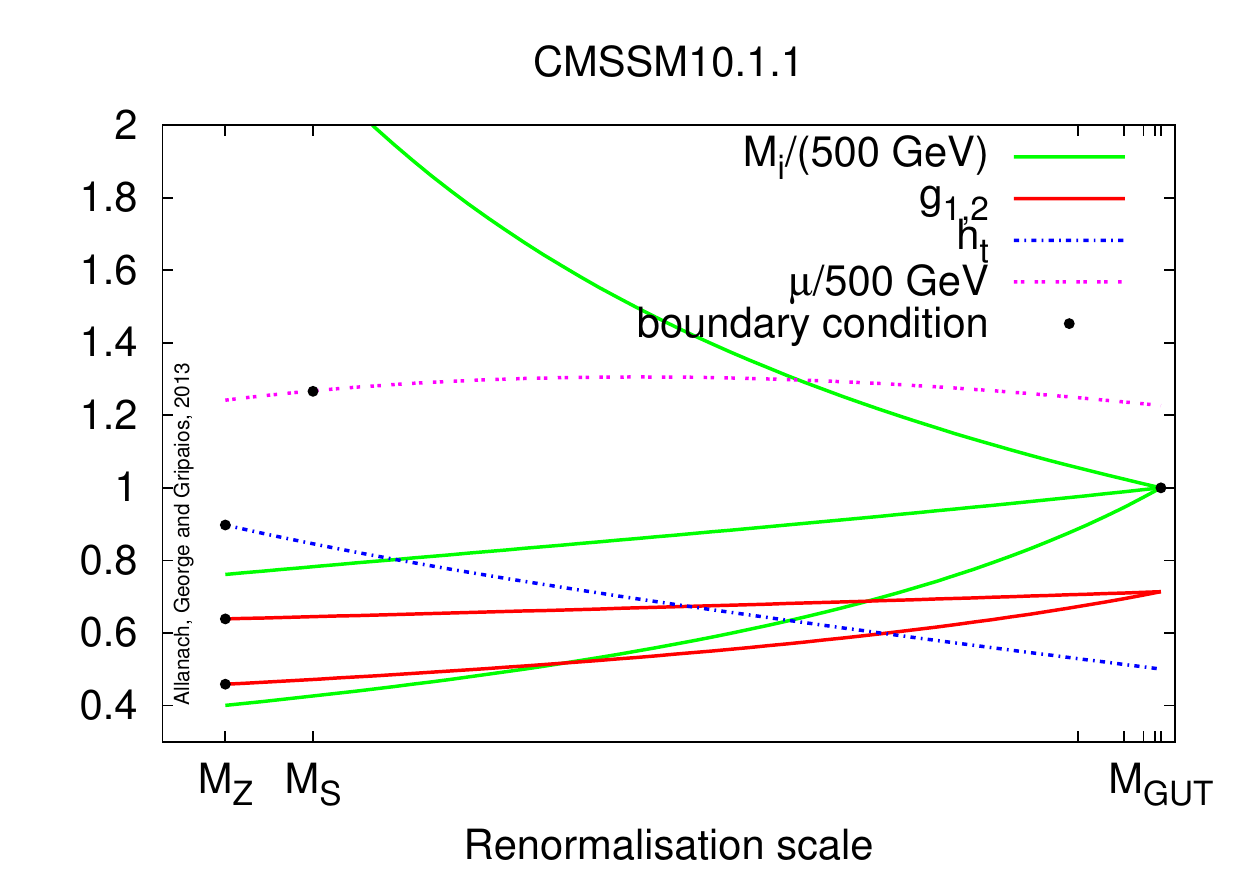}
\caption{Boundary conditions and renormalisation group flow at
  CMSSM10.1.1~\cite{AbdusSalam:2011fc}. The abscissa is a logarithmic
  scale. 
\label{fig:bcs}}
\end{center}
\end{figure}

Before giving the boundary conditions in some detail, we provide a
rough sketch of them (and of 
a resulting RG flow) in Fig.~\ref{fig:bcs} for a certain CMSSM
point. 
Black dots show the boundary conditions: at the low scale, we have
boundary conditions on gauge and Yukawa couplings coming from experimental
data. In the Figure, $\msusy$ is the scale at which
  superparticle masses are calculated and the Higgs potential is minimised. 
Thus, at $\msusy$, we have a boundary condition on $\mu(\msusy)$ coming
from the Higgs potential minimisation. At the high renormalisation
scale, we have a boundary condition on the SUSY breaking terms (we have
illustrated this with the gaugino masses $M_i$). The high scale itself,
$\mgut$, is usually set to be the scale $Q$ at which $g_1(Q)=g_2(Q)$,
providing another boundary condition ($g_3(\mgut)$ is close to these, but
not exactly 
unified in the MSSM; this small discrepancy can easily be explained by GUT
threshold corrections~\cite{Yamada,Altarelli:2000fu}). 
The boundary conditions are linked by the
RG flow, a set of differential equations in these variables.

The RG equations of the MSSM are non-linear,
coupled, homogeneous first order equations. 
For example, at one-loop order we have equations governing the evolution of
the two Higgs soft SUSY breaking mass parameters $m_{H_1}$ and
$m_{H_2}$ of the form \cite{Martin:1993zk}
 \begin{eqnarray}
16 \pi^2\frac{\partial m_{H_2}^2}{d t} &=& 
6 \left[ (m_{H_2}^2 + m_{{\tilde Q}_3}^2 +
  m_{{\tilde u}_3}^2 + A_t^2)  h_t^2  \right] - 6 g_2^2 M_2^2 - \frac{6}{5}
g_1^2 M_1^2 + \frac{3}{5}
g_1^2 \left(m_{H_2}^2-m_{H_1}^2 + \right. \nonumber \\
&&\left. \mbox{Tr}[m_{\tilde Q}^2 - m_{\tilde L}^2 - 2 
m_{\tilde u}^2 + m_{\tilde d}^2 + m_{\tilde e}^2] \right), \label{rgesA} \\
16 \pi^2\frac{\partial m_{H_1}^2}{d t} &=& 
6 \left[ (m_{H_1}^2 + m_{{\tilde Q}_3}^2 +
  m_{{\tilde d}_3}^2 + A_b^2)  h_b^2  \right] - 6 g_2^2 M_2^2 - \frac{6}{5}
g_1^2 M_1^2 -
\frac{3}{5}
g_1^2 \left(m_{H_2}^2-m_{H_1}^2 + \right. \nonumber \\
&&\left. \mbox{Tr}[m_{\tilde Q}^2 - m_{\tilde L}^2 - 2 
m_{\tilde u}^2 + m_{\tilde d}^2 + m_{\tilde e}^2] \right), \label{rges}
\end{eqnarray}
where $t=\ln Q$. There are other similar equations for each of the quantities
on the 
right-hand side of Eq.~\eqref{rges}.

We  specify our boundary conditions explicitly below for: the Yukawa
couplings of the top, bottom, and tau, {\em viz.} $h_t$, $h_b$, and
$h_\tau$, respectively;
the 3 MSSM gauge couplings, $g_i$, $i \in \{1,2,3\}$; the various SUSY
breaking scalar 
masses, $m_{\varphi}$; the gaugino masses, $M_i$, $i \in \{1,2,3\}$; and for
the SUSY breaking 
trilinear scalar couplings for top, bottom, and tau, $A_t, A_b$ and $A_\tau$. 
Other boundary conditions include those on the
parameter $\mu$ appearing in the superpotential\footnote{The circumflex
  indicates a superfield.} $W \supset \mu \hat H_1
\hat H_2$, and on $m_3^2$, that mixes the two Higgs doublets in the potential $V \supset m_3^2
H_2 H_1$. These last two boundary
conditions come from the minimisation of the Higgs potential with
respect to the neutral components of  
$H_1$ and 
$H_2$.

In all,
we have the boundary conditions
\begin{eqnarray}
\tan \beta(M_Z) &=& \tan \beta (\mbox{input}) \label{tanb} \\
h_t({M_Z}) &=& \frac{m_t(M_Z) \sqrt{2}}{v(M_Z) \sin \beta}, \qquad
h_{b,\tau}(M_Z) = \frac{m_{b,\tau}(M_Z) \sqrt{2}}{v(M_Z) \cos
  \beta}, \label{Yukawas}\\
v(M_Z) &=& 2 \sqrt{\frac{{M_Z^2(\mbox{exp}) + \Pi_{ZZ}^T(M_Z)}}{{\frac{3}{5}
    g_1^2(M_Z) + 
    g_2^2(M_Z)}}} \label{vev}\\ 
g_1(M_Z)&=&g_1(\mbox{exp}), \qquad g_2(M_Z) = g_2(\mbox{exp}), \qquad g_3(M_Z) =
g_3(\mbox{exp}). \label{gaugeCouplings} \\
\msusy &=& \sqrt{m_{{\tilde t}_1}(\msusy) m_{{\tilde
      t}_2}(\msusy)} \label{msusy} \\
\mu^2(\msusy) &=&  
\frac{m_{\bar{H}_1}^2(\msusy) -  m_{\bar{H}_2}^2(\msusy) \tan^2
  \beta(\msusy)}{\tan^2 \beta(\msusy) - 1} 
- \frac{1}{2} M_Z^2(\msusy)
\label{mucond} \\
m_3^2(\msusy)&=&\frac{\sin {2\beta(\msusy)}}{2} \left( m_{\bar{H}_1}^2(\msusy) + m_{\bar{H}_2}^2(\msusy) + 2 \mu^2(\msusy)
\right) \label{Bcond} \\
g_1(\mgut) &=& g_2(\mgut) \label{unifc} \\
M_1(\mgut) &=& M_2(\mgut) = M_3(\mgut) = M_{1/2} \label{mhalf} \\
m_{\tilde u}^2(\mgut) &=& m_{\tilde d}^2(\mgut) = m_{\tilde e}^2(\mgut) = m_{\tilde L}^2(\mgut) =
m_{\tilde Q}^2(\mgut) = m_0^2 I_3 \label{sparticle}
\\ m_{H_1}^2(\mgut) &=&
m_{H_2}^2(\mgut) =m_0^2 \label{mzero} \\ 
A_{\tilde u}(\mgut) &=& A_0 I_3, \qquad
A_{\tilde d}(\mgut) = A_0 I_3, \qquad
A_{\tilde e}(\mgut) = A_0 I_3. \label{trilinears}
\end{eqnarray}
The running parameters in Eqs.~\eqref{tanb}-\eqref{trilinears} are in the modified
dimensional reduction
(DRED) scheme~\cite{Capper:1979ns}. The `(exp)' denotes that the value derives from experimental
data. 
We have labelled the input
parameter $\tan \beta$  
as $\tan \beta(\mbox{input})$. The parameters
$m_{b,t,\tau}(M_Z)$ and $g_{1,2,3}(M_Z)$ are obtained from experimental
data, subtracting loops due 
to sparticles and Standard Model particles. The Standard Model electroweak
gauge couplings  
$g_1(\mbox{exp})$ and $g_2(\mbox{exp})$ are fixed by the fine structure
constant $\alpha$ and the Fermi constant, $G_F$. The values of
$g_i(\mbox{exp})$ are corrected by 
one-loop corrections involving sparticles.
The parameter $v(M_Z)\approx 246$ GeV denotes
$\sqrt{v_1^2(M_Z)+v_2^2(M_Z)}$, where $v_1$ and $v_2$ are the vacuum
expectation values of the neutral components of the Higgs doublets $H_1$ and
$H_2$, respectively. 
The modified DRED $Z^0$ boson mass squared is fixed by $M_Z^2(\msusy)=v^2(\msusy)
\left(\frac{3}{5} g_1^2(\msusy) + g_2^2(\msusy)\right)/4$.
$m_{\bar{H}_i}^2=m_{H_i}^2-t_i/v_i$  are fixed by the soft SUSY
breaking mass parameters for the Higgs fields $m_{H_i}^2$, 
$i \in \{1,2\}$, as well as by the tadpole
contributions $t_i$ coming from loops. 
These tadpole contributions have terms linear in $\mu(\msusy)$ as well as
terms that are logarithmic in it, and so Eq.~\eqref{mucond} is not a
simple quadratic equation for $\mu(\msusy)$.
$\Pi_{ZZ}^T(M_Z)$ is the MSSM self-energy correction to the $Z^0$ boson mass
which can be found in Ref.~\cite{Pierce:1996zz} and
$I_3$ is a 3$\times$3
matrix in family space. 
For further details, see the {\tt
  SOFTSUSY}~manual~\cite{Allanach:2001kg}. 

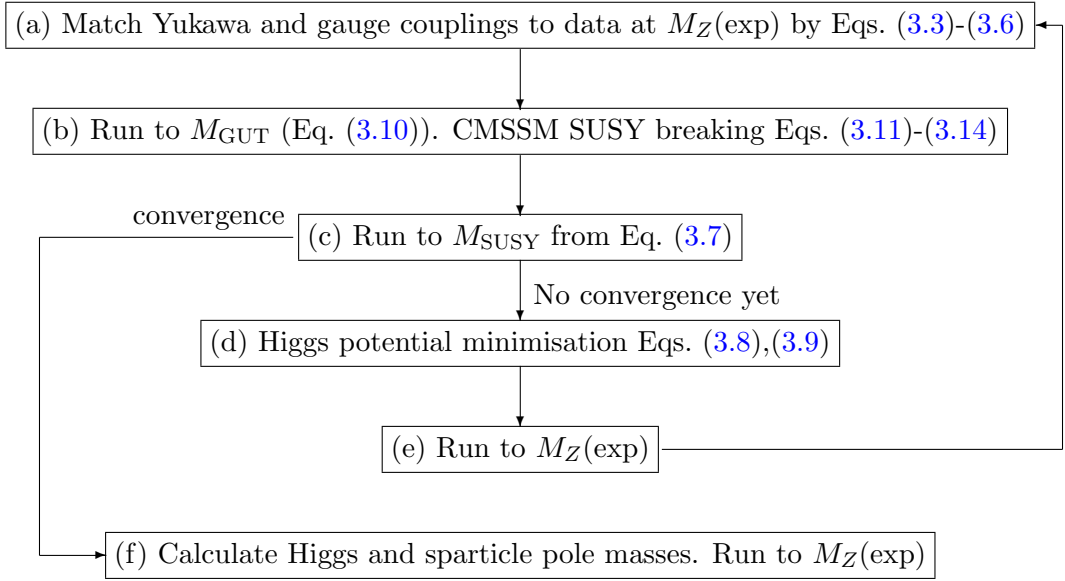
\begin{figure}\begin{center}
\label{fig:algorithm}
\begin{picture}(323,210)
\put(10,0){\makebox(280,10)[c]{\fbox{(f) Calculate Higgs and
      sparticle pole masses. Run to $M_Z(\mbox{exp})$}}}
%\put(150,36.5){\vector(0,-1){23}}
\put(10,40){\makebox(280,10)[c]{\fbox{(e) Run to $M_Z(\mbox{exp})$}}}
\put(150,76.5){\vector(0,-1){23}}
\put(10,160){\makebox(280,10)[c]{\fbox{(b) Run to $\mgut$
      (Eq.~\eqref{unifc}). CMSSM SUSY breaking
      Eqs.~\eqref{mhalf}-\eqref{trilinears}}}} 
\put(150,116.5){\vector(0,-1){23}}
\put(10,80){\makebox(280,10)[c]{\fbox{(d) Higgs potential minimisation Eqs.~\eqref{mucond},\eqref{Bcond}}}}
\put(150,156){\vector(0,-1){23}}
\put(10,120){\makebox(280,10)[c]{\fbox{(c) Run to $\msusy$ from
      Eq.~\eqref{msusy}}}} 
\put(65,125){\line(-1,0){95}}
\put(-30,125){\line(0,-1){120}}
\put(-30,5){\vector(1,0){25}}
\put(5,130){convergence}
\put(155,100){No convergence yet}
\put(150,196){\vector(0,-1){23}}
\put(10,200){\makebox(280,10)[c]{\fbox{(a)
Match Yukawa and gauge couplings to data at $M_Z(\mbox{exp})$ by Eqs.~\eqref{tanb}-\eqref{gaugeCouplings}}}}  
\put(203,45){\line(1,0){150}}
\put(353,45){\line(0,1){160}}
\put(353,205){\vector(-1,0){10}}
\end{picture}
\caption{Iterative algorithm used to calculate the SUSY spectrum. Each step
(represented by a box) is described in more detail in the text. The initial step is the
uppermost one.}\end{center}\end{figure}

Spectrum calculators for the MSSM that
are currently in the public domain, namely {\tt
  ISASUSY}~\cite{Baer:1993ae}, {\tt 
  SOFTSUSY}~\cite{Allanach:2001kg}, {\tt   SPheno}~\cite{Porod:2003um}, {\tt
  SUSEFLAV}~\cite{Chowdhury:2011zr} and {\tt SUSPECT}~\cite{Djouadi:2002ze},
solve 
Eqs.~\eqref{tanb}-\eqref{trilinears} (or equations very similar to them)
and thus find `the'
RG flow by fixed point iteration. 
The particular algorithm used by {\tt SOFTSUSY}, for example, is shown in
Fig.~\ref{fig:algorithm}. 
Fixed point iteration can only find at most one solution, for a given starting
point: guesses for MSSM parameters 
are initially input to step (a), and the algorithm is applied. If a
self-consistent solution to the whole system of boundary conditions and RG
flow is found, the parameters at step (c) remain approximately the same on
successive iterations, and the iterative loop is exited, returning a single
solution. If we are to find multiple solutions, this iterative algorithm must
be modified. 

\subsection{Finding some multiple solutions}
In order to exhibit multiple solutions, we follow the aforementioned
prescription of 
changing the boundary conditions slightly. We leave all of the boundary
conditions described above unchanged, except for the one for
$\mu({\msusy})$, which we allow to be an input parameter instead. Thus
we do not apply Eq.~\eqref{mucond}: instead, we turn it into a prediction
$M_Z(\mbox{pred})$ for the $Z^0$ boson pole mass:
\begin{eqnarray}
M_Z^2(\mbox{pred}) &=&
2 \left( \frac{m_{\bar{H}_1}^2(\msusy) -  m_{\bar{H}_2}^2(\msusy) \tan^2
  \beta(\msusy)}{\tan^2 \beta(\msusy) - 1} -  \mu^2(\msusy)
\right)\nonumber \\ && + \Pi_{ZZ}^T(\msusy),
\label{mzpred} 
\end{eqnarray}
where in practice, we use $\Pi_{ZZ}^T(\msusy) = M_Z^2(\mbox{exp}) -
M_Z^2(\msusy)$. 
When Eq.~\eqref{mzpred} agrees
with the experimentally determined central value, $M_Z(\mbox{exp})=91.1887$
GeV, we 
have a consistent solution of the system of boundary conditions and
renormalisation group equations. Thus, in the algorithm, we supplant
Eq.~\eqref{mucond} by Eq.~\eqref{mzpred} in step (d). This is much the
same approach as the one we took in the BKT toy example above, in that we have
relaxed a 
boundary condition, {\em viz.}\/ Eq.~\eqref{mucond}, and scanned over a
coupling, {\em viz.}\/ $\mu({\msusy})$, that appeared in it. Our
algorithm can still find at most one solution for each
value of $\mu({\msusy})$, but more than one value of $\mu({\msusy})$ might
satisfy the original
boundary conditions.

% START CHANGES
We emphasise that, while $\mu(\msusy)$ is scanned, no other parameters are
changed by hand.  A change in $\mu(\msusy)$ changes the neutralino, chargino
and third family squark masses.  Thus, the SUSY radiative corrections to the
top mass will change, and therefore $h_t$ via Eq.~\eqref{Yukawas}.  $h_t$, in
turn, strongly affects the renormalisation of $M_{H_i}^2$, thus $M_{H_i}^2(\msusy)$
may change, even though $M_{H_i}^2(\mgut)$ remains fixed by the selected point
in the CMSSM parameter space.
% END CHANGES

\begin{figure}\begin{center}
\begin{picture}(300, 280)(0,0)
\put(-50,280){\includegraphics[width=280pt, angle=270]{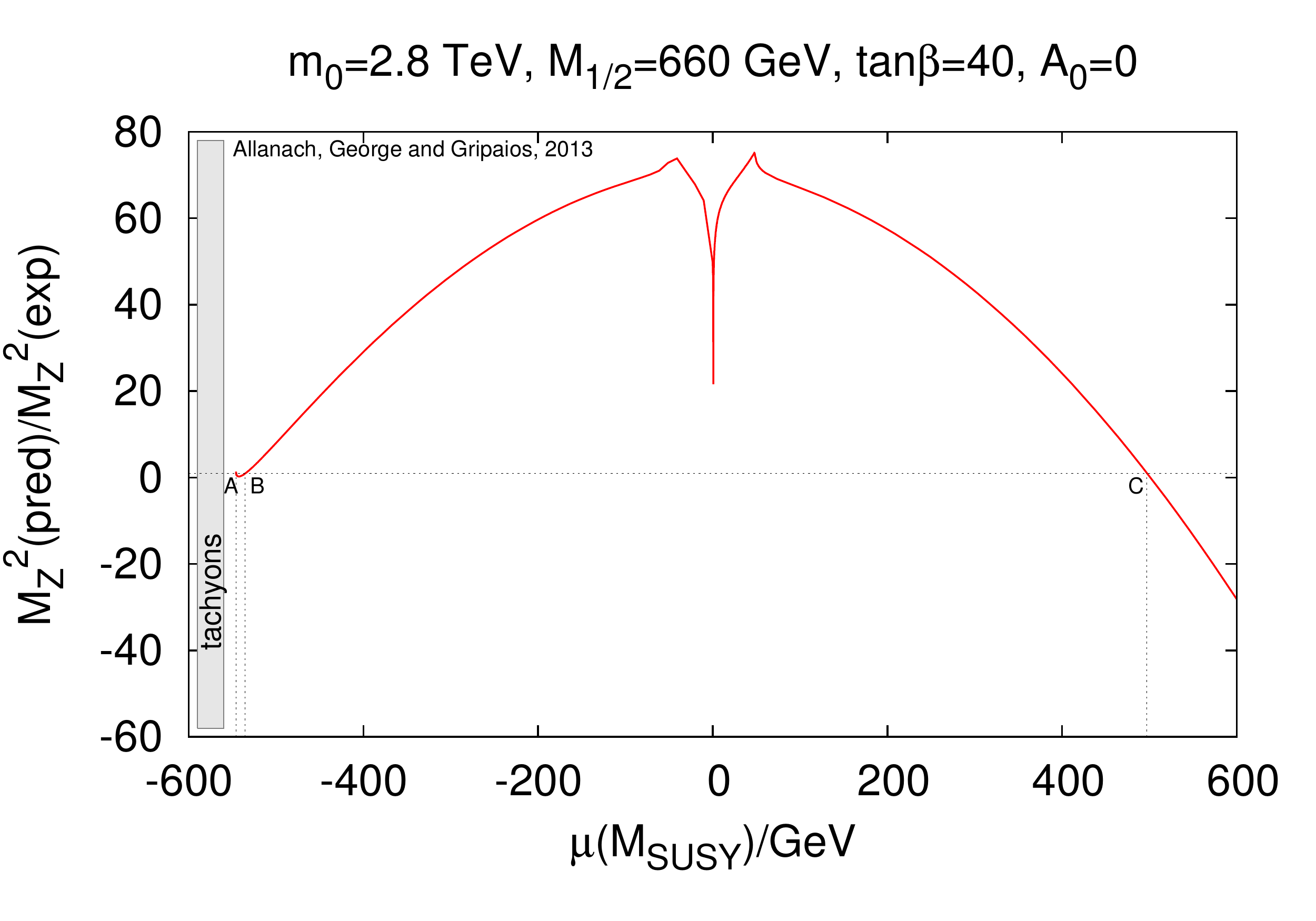}}
\put(30,135){\includegraphics[width=80pt, angle=270]{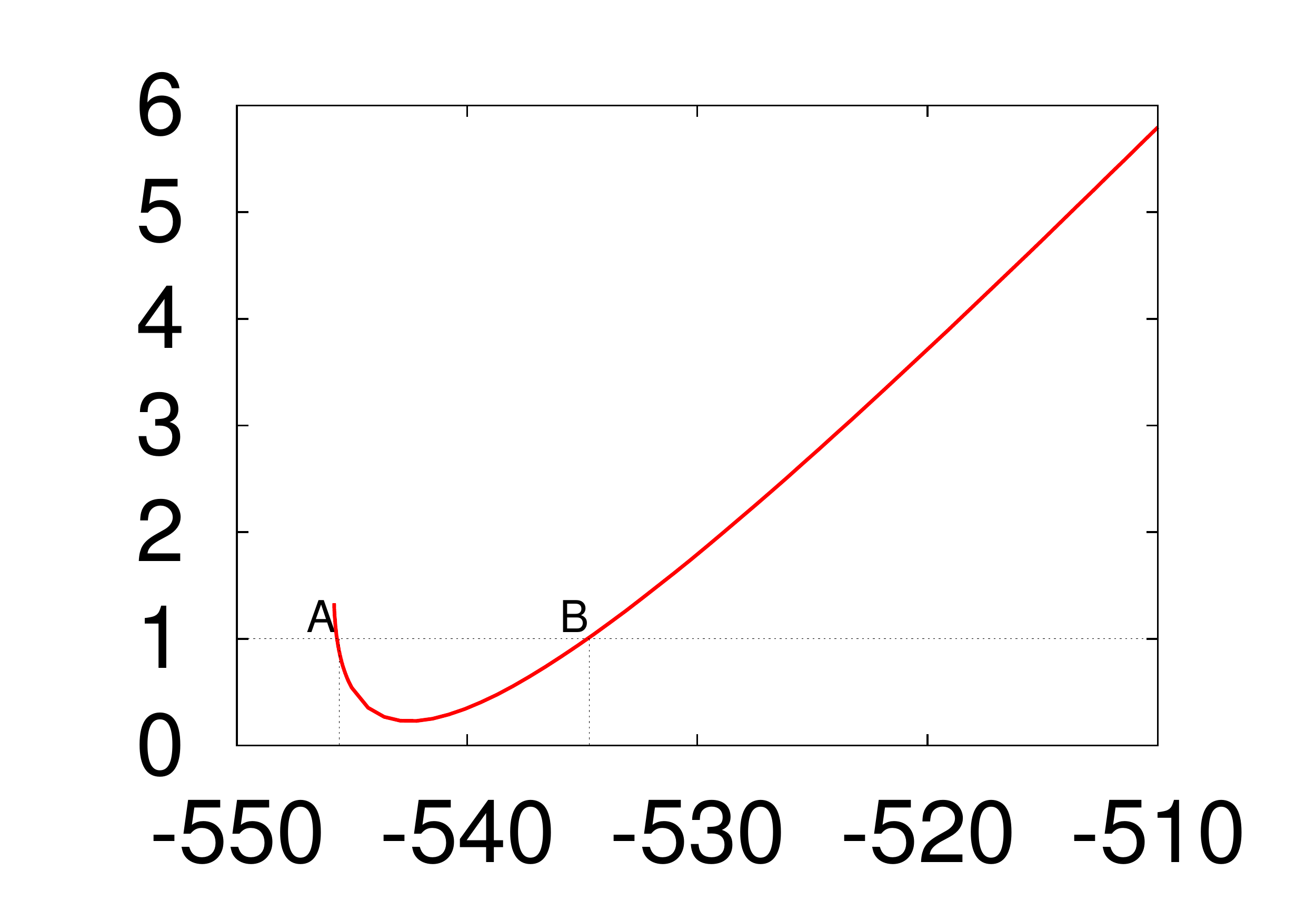}}
\end{picture}
\caption{Multiple solutions in the CMSSM at $m_0=2800$ GeV, $M_{1/2}=660$ GeV, 
$A_0=0$ and $\tan   \beta=40$. Along the curve, all
  boundary conditions are applied except for the Higgs potential minimisation
  condition for $\mu(\msusy)$, Eq.~\eqref{mucond}. 
  Where the curve intersects
  the line $M_Z^2(\mbox{pred})/M_Z^2(\mbox{exp})=1$
  (marked as A, B and C), there are solutions that
  are consistent with all of the boundary conditions. The insert shows a
  zoomed region of the plot.
\label{fig:multiPheno}}
\end{center}
\end{figure}

We thus use a modified version of {\tt SOFTSUSY3.3.7} with the altered
algorithm, and Standard Model 
input parameters as listed in Appendix~\ref{sec:inputs}.
We show $M_Z(\mbox{pred})$ as a function of
$\mu(\msusy)$ for one such CMSSM point in
Fig.~\ref{fig:multiPheno}. For $\mu<-550$ GeV, there is no physical solution
due to the pseudoscalar $A^0$ being tachyonic, signalling that the desired
electroweak minimum 
is not 
present. 
In the Figure, we see  the usual two solutions (one for $\mu>0$ at
point B
and one for
$\mu<0$ at point C, both of which {\tt SOFTSUSY3.3.7} finds) plus an additional solution
(at the point A) for $\mu<0$. Along the curve, $M_X$, gauge
 and Yukawa couplings, and Higgs soft mass parameters vary, indeed all MSSM
 parameters that are not boundary conditions, vary. 
Between $\mu(\msusy)=-548$ GeV
and $\mu(\msusy)=-530$ GeV, each term on the right-hand side of
Eq.~\eqref{mzpred} is monotonic, and so it is only their combination that
renders the minimum. In particular, it is the combination of the terms within
the large brackets, since $M_Z^2(\msusy)$ has only a very tiny dependence upon
$\mu(\msusy)$ within its range.
We have strong evidence to suggest that there are no other solutions at larger
values of $|\mu(\msusy)|$: there, the second term $-2\mu^2(\msusy)$ in
Eq.~\eqref{mzpred} dominates, and $M_Z^2(\mbox{pred})$ is negative. 

Out-of-the box~{\tt SOFTSUSY3.3.7} will not find solution A, even if the
starting guess for the iteration is changed. This is because the solution is
repulsive if one uses the unmodified algorithm depicted in
Fig.~\ref{fig:algorithm}. We have checked this by inputting solution A into
the usual {\tt SOFTSUSY3.3.7} 
algorithm and performing some iterations: the algorithm then
converges on to solution B. We may consider the usual {\tt SOFTSUSY3.3.7}
algorithm to be an 
iterative fixed point algorithm in $\mu$: then, for a solution, we have
$\mu_0(\msusy)=f(\mu_0(\msusy))$, where $f$ is the function that performs one
iteration i.e.\ the ordered steps (d), (e), (a), (b) and (c) in
Fig.~\ref{fig:algorithm}. 
For some value of $\mu=\mu_0(\msusy)$ corresponding to a solution, 
in a neighbourhood around the point $\mu_0$ 
the fixed
point iteration algorithm is {\em stable} if $|d f(\mu(\msusy))/d
\mu(\msusy)|_{\mu=\mu_0} < 1$ and {\em unstable} if $|d f(\mu(\msusy))/d
\mu(\msusy)|_{\mu=\mu_0} > 1$. We calculate $f(\mu_0(\msusy))$ numerically by
first performing the {\em modified} iterative algorithm to calculate $\mu_0$,
then running one 
standard {\tt SOFTSUSY} iteration upon the result. 
This allows a numerical determination: indeed $|d f(\mu(\msusy))/d
\mu(\msusy)|=7.4,0.3,0.3$ for solutions A, B and C, respectively,
demonstrating again that A is unstable with respect to fixed point iteration,
whereas B and C are stable. 

The multiple solutions are not due to non-linearities in Eq.~\eqref{mzpred}
alone.  
We have shown this by
taking each of our solutions in turn (say, A) and then scanning over
$\mu(\msusy)$ while 
{\em not}\/ performing the RG flow and not applying the other boundary
conditions, 
calculating $M_Z^2(\mbox{pred})/M_Z^2(\mbox{exp})$. This system is
observed to have only one solution for a given sign of $\mu$. 
This is in contrast to the multiple solutions found in a model with an extra
constraint (mSUGRA) by Drees and
Nojiri~\cite{Drees:1991ab}, where a single boundary condition (namely the one
for $\tan \beta$ derived from 
Eq.~\ref{Bcond}) displayed
multiple solutions in some regions of parameter space. 

Fig.~\ref{fig:multiPheno} displays some non-smooth kinks at $\mu \approx \pm
50$ GeV: these occur when $|M_{\chi_1^0}|+|M_{\chi_2^0}|=M_Z$. For $|\mu| > 50$
GeV, a non-smooth piece is introduced in the one-loop 
self-energy of the $Z^0$, $\Pi_{ZZ}^T(\msusy)$ as the particles in the loop
are no longer  on-shell. The 
inverted spike at $\mu(\msusy) \approx 6$ GeV  also is present for 
other values of $m_0$, $M_{1/2}$, $A_0$ and $\tan \beta$, and can lead to
additional solutions, if it is deep enough. 
At $\mu(\msusy) \approx 6$ GeV, the charginos are approaching zero mass,
changing the value of the 
running values of the MSSM electromagnetic coupling as extracted from
data. This changes the electroweak couplings, which changes $\mgut$,
significantly changing in turn $m_{H_2}^2(\msusy)$ in Eq.~\eqref{mzpred},
which dominates the prediction for $M_Z(\mbox{pred})$.

\begin{table}
\begin{center}
\begin{tabular}{|c|ccc|}\hline
quantity & solution A & solution B & solution C \\ 
\hline 
\hline 
$\mu(\msusy)$/GeV & -545 & -535& 497 \\ 
\hline 
$M_{\chi_1^0}$/GeV    & 282 & 282 & 281 \\
$M_{\chi_2^0}$/GeV    & 502 & 497& 471 \\
$M_{\chi_3^0}$/GeV    & 558 & 548 & 510 \\
$M_{\chi_4^0}$/GeV    & 610 & 605 & 593 \\
$M_{\chi_1^\pm}$/GeV & 503& 497& 470 \\
$M_{\chi_2^\pm}$/GeV & 609 & 604 & 592 \\
$m_{\tilde g}$/GeV    & 1612 & 1612 & 1612  \\ \hline
$m_3^2 (\msusy)/10^5$ GeV$^2$ & 0.800 & 0.809 & 1.07 \\
$m_{H_2}^2 (\msusy)/10^5$ GeV$^2$ & -1.94 & -1.83& -1.42 \\
$h_t(\msusy)$ & 0.840& 0.839 & 0.836 \\
$A_t(\msusy)$/GeV & -1056&-1057 & -1064 \\
$M_X/10^{16}$ GeV & 1.94 & 1.93& 1.89 \\
$g_1(M_Z)$ & 0.460 & 0.470& 0.456 \\
$g_2(M_Z)$ & 0.634 & 0.640& 0.633 \\ 
\hline\end{tabular}
\caption{\label{tab:spec} Differences in CMSSM parameters and spectra for the
  multiple 
  solutions of the parameter point $m_0=2.8$ TeV, $M_{1/2}=660$ GeV, $\tan
  \beta=40$ and $A_0=0$ displayed in Fig.~\protect\ref{fig:multiPheno}. 
  The solutions are found by scanning $\mu(\msusy)$ and then the rest of the
  quantities are determined by the iterative algorithm.
  We display here some masses and parameters of interest for the 3 solutions
  that predict the correct value of $M_Z$.
  Above the central horizontal line, all masses are pole masses, whereas below
  the line, all quantities are evaluated in the modified DRED scheme.
  Out-of-the-box {\tt SOFTSUSY3.3.7} finds solutions B and C.}
\end{center}
\end{table}
We display the respective spectra of solutions A,B and C in
Table~\ref{tab:spec}.  
The spectra show some notable
differences, illustrating the fact that the solutions are physically
different, leading to the possibility of their discrimination by collider
measurements. Masses whose tree-level values depend upon the value of
$\mu$, such as the heavier neutralino and chargino masses, 
show the largest differences. Other
sparticle and Higgs masses do have small per-mille level differences. 
We also see some differences in the modified
DRED scheme running parameters between the solutions. 
We have found
other points in CMSSM parameter space with several solutions where some
sparticle masses
differ by hundreds of GeV, but the additional solutions had sparticles lighter than
$M_Z/2$ and were
obviously phenomenologically excluded by LEP, which saw no significant
evidence for sparticles in millions of $Z^0$ decays. 

\begin{figure}\begin{center}
\includegraphics[width=0.8\textwidth]{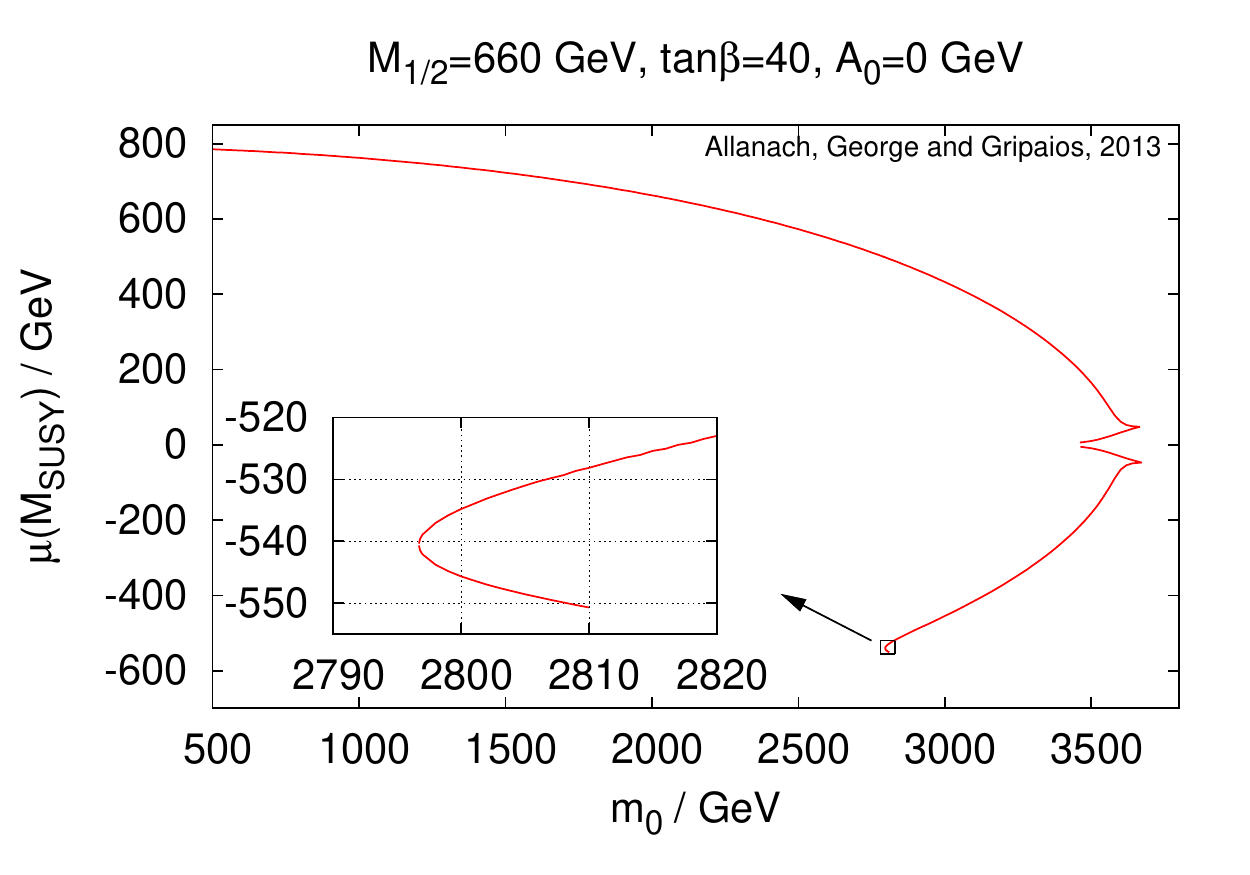}
\caption{Multiple branches of solutions in the CMSSM at $M_{1/2}=660$~GeV,
  $A_0=0$~GeV and $\tan\beta=40$.  Each point on the curve is a
  solution consistent with all boundary conditions.  The value of $m_0$ ranges
  as per the horizontal axis, and corresponds to a horizontal slice through
  Fig.~\ref{fig:multiSolnsScan}.
  The curve exhibits a single solution for $m_0<2796.7$~GeV and
  multiple solutions (3 then 2 then 4) as $m_0$ increases above
  this value.
\label{fig:multiPheno-muvsm0}}
\end{center}
\end{figure}
\begin{figure}\begin{center}
\includegraphics[width=0.8\textwidth]{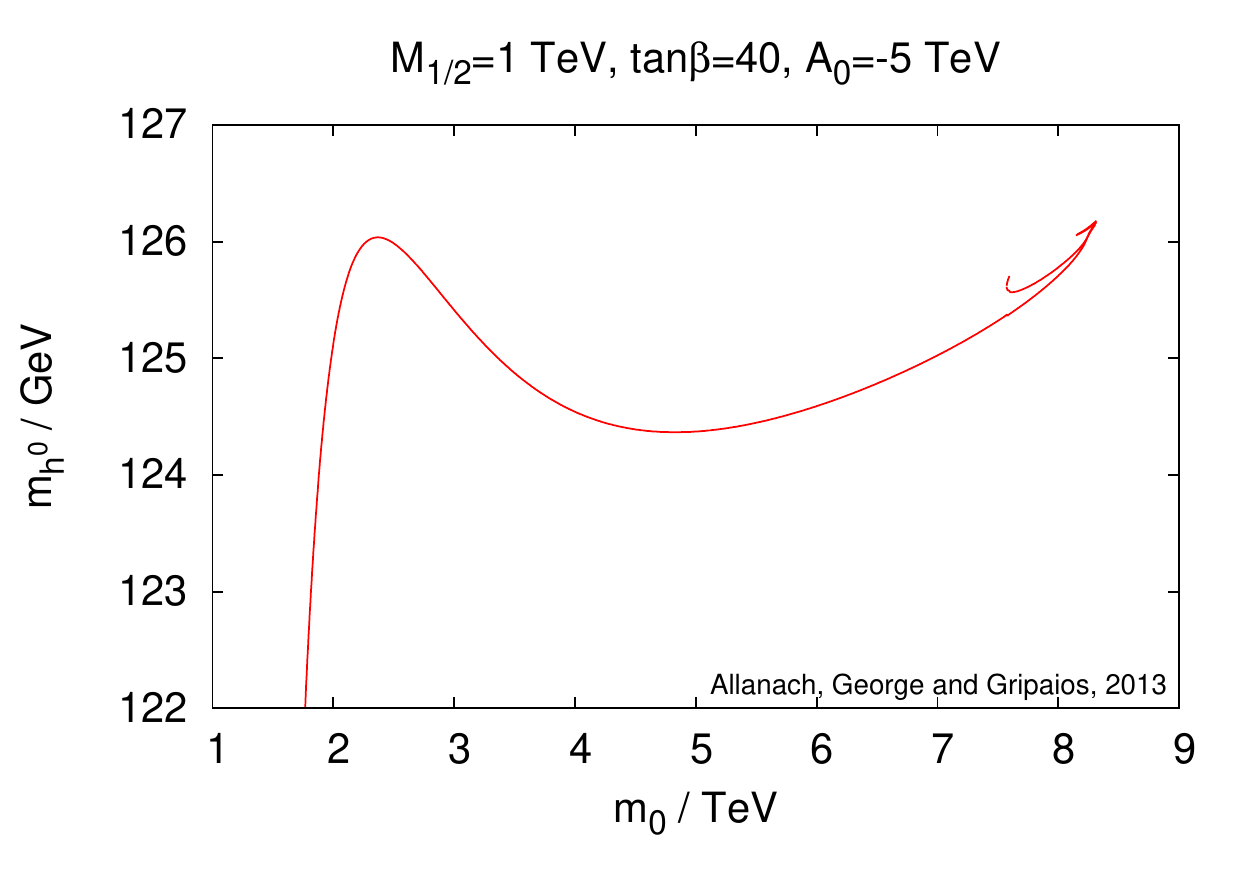}
\caption{Multiple branches of solutions in the CMSSM at $M_{1/2}=1$~TeV,
  $A_0=-5$~TeV and $\tan\beta=40$. The value of $m_0$ ranges as per the horizontal
  axis.  We plot here the predicted values of the Higgs mass, $m_{h^0}$.  There are
  3 solutions in the range $7572\;\text{GeV} \leq m_0 \leq 7595\;\text{GeV}$
  which are consistent with recent LHC measurements of a Higgs boson mass.
\label{fig:multiPheno-mh0vsm0}}
\end{center}
\end{figure}
The CMSSM parameter point at which we have found our multiple solutions A, B,
C is by
no means alone. In Fig.~\ref{fig:multiPheno-muvsm0}, we scan in $m_0$ as well as
$\mu(\msusy)$ in order to show the appearance and disappearance of various
branches of the solutions. We see the appearance of multiple solutions for 
$m_0>2796.7$ GeV, which then collapse to the usual two solutions, one for
$\mu>0$ and one for $\mu<0$, when $m_0=2810-3400$ GeV. For $m_0=3400-3700$ GeV
though, there are four solutions.  In Fig.~\ref{fig:multiPheno-mh0vsm0} we exhibit
another point in parameter space that has multiple branches of solutions which
predict a CP-even lightest Higgs mass consistent with recent LHC measurements of
a Higgs boson~\cite{ATLAS:2012oga,Chatrchyan:2012ufa}.  For
$m_0=7572-7595$ GeV\footnote{As is well known, getting $m_h \sim 125$
  GeV in the MSSM in general requires unnaturally large stop quark masses, and
  hence large $m_0$ in the CMSSM.}
there are 3 solutions, 2 with $\mu(\msusy)<0$, which have $m_{h^0}$ in
the range $125.4$ GeV to $125.7$ GeV.

\begin{figure}\begin{center}
\includegraphics[angle=90, width=0.6\textwidth,angle=270]{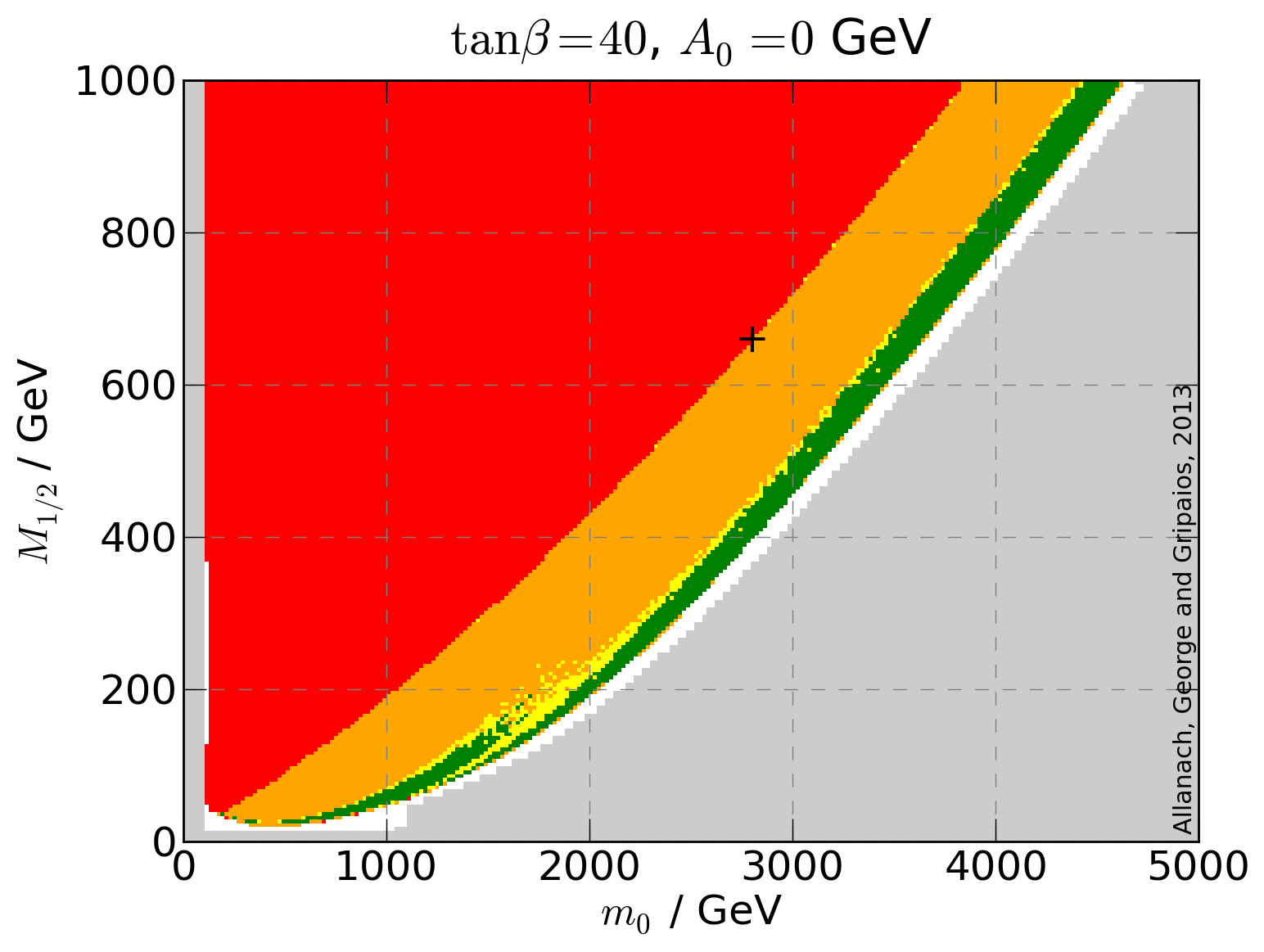}
\caption{Parameter scan of the CMSSM, with colour indicating number of solutions
    at a given point: white, red, orange, yellow and green being 0, 1, 2, 3 and 4
    solutions respectively.  The red region at the top-left has 1 solution
    for $\mu(\msusy)>0$ and no solutions for $\mu(\msusy)<0$.  The orange
    region to the right of this red region has 2 solutions, 1 for each sign
    of $\mu(\msusy)$.  Further to the right are regions with 3 and 4
    solutions.
    The `+' marks the position of the point detailed in Table~\ref{tab:spec},
    and lies on a thin region, difficult to see in between the red and orange
    regions, which possesses 3 solutions.
    Phenomenological constraints have not been applied.
\label{fig:multiSolnsScan}}
\end{center}
\end{figure}
We now wish to investigate how prevalent in CMSSM parameter space the multiple
solutions are. In Fig.~\ref{fig:multiSolnsScan}, we fix $\tan\beta$ and $A_0$, and
scan over $m_0$ and $M_{1/2}$, showing by colour how many solutions
(with either sign of $\mu$) we are able to find at each point.  White regions correspond
to no solution (because, for example, there is no electroweak symmetry breaking),
and red, orange, yellow and green correspond to 1, 2, 3, and 4 solutions,
respectively.  Grey regions have not been scanned, but the large grey area on the right
of the plot is in the usual region of unsuccessful electroweak symmetry
breaking, and we expect
0 solutions.  The large red region has a single solution for $\mu>0$; in this region
there is no solution for $\mu<0$ due to tachyonic $A_0$.  The orange region has the
usual 2 solutions, one for each sign of $\mu$.  Beyond this, near the boundary
of electroweak symmetry breaking, multiple solutions for both signs of $\mu$ are prevalent.
The solutions found near the edge of successful electroweak symmetry breaking correspond
to the spiky apex of the curve in
Fig.~\ref{fig:multiPheno} shifting down to intersect the horizontal dotted line in 4
distinct places, corresponding to 4 correct predictions for $M_Z$.

Fig.~\ref{fig:multiPheno-muvsm0} shows the value of $\mu$ for a slice through
Fig.~\ref{fig:multiSolnsScan} at $M_{1/2}=660$ GeV.  For low $m_0$ we are in the
red region and there is only 1 solution.  As we increase $m_0$, just before entering
the region with 2 solutions, we hit a small patch of parameter space which exhibits
3 solutions, 2 of which are for $\mu<0$.  The points in parameter space where this
occurs are indicated (for those with keen eyesight) in Fig.~\ref{fig:multiSolnsScan} by the isolated yellow dots between the red and orange
regions.  We have verified that these regions of 3 multiple solutions are not confined
to single points in parameter space, but rather occupy a finite area, as shown
by the inset in Fig.~\ref{fig:multiPheno-muvsm0}, and as shown by the small region
$m_0=7572-7595$ GeV in Fig.~\ref{fig:multiPheno-mh0vsm0}.
It is expected that this area runs the length 
of the boundary between the red and orange regions.  Whilst most of the multiple
solutions are fairly near the boundary of electroweak symmetry breaking, there is
nevertheless a significant volume of parameter space where the multiple solutions play
a r\^{o}le, and they should not be ignored in phenomenological analyses.

We have performed further scans of parameter space at selected values of
$\tan\beta$ and $A_0$, and counted solutions in the $m_0$-$M_{1/2}$ plane.
The qualitative features of these other slices through the parameter space
of the CMSSM are similar to Fig.~\ref{fig:multiSolnsScan}, including the
existence of a region with three solutions between the red and orange regions.
One occasionally finds small areas, close to the edge of successful electroweak
symmetry breaking, that have six solutions, two for $\mu<0$ and four for
$\mu>0$. 
For example, $\tan\beta=20$, $A_0=-1000$ GeV, $M_{1/2}=170$ GeV and $m_0=2840$
GeV is a point with six such solutions. The regions with three and four
solutions occur at larger values of $m_0$ close to the region of electroweak
symmetry breaking. While the position of the no electroweak symmetry breaking  boundary is extremely sensitive to
the top quark mass $m_t$, so that uncertainties on it remain
high~\cite{Allanach:2000ii,Allanach:2012qd}, we have checked that the
additional solutions remain (but move in $m_0$) when we change $m_t$.

%%%%%%%%%%%%%%%%%%%%%%%%%%%%%%%%%%%%%%%%%%%%%%%%%%%%%%%%%%%%%
\section{Discussion \label{sec:disc}}
%%%%%%%%%%%%%%%%%%%%%%%%%%%%%%%%%%%%%%%%%%%%%%%%%%%%%%%%%%%%%

Quantum field theories with multiple parameters that have boundary
conditions imposed at multiple renormalisation scales may admit several
physically distinct solutions for the same boundary conditions.  Each separate
solution corresponds to a different RG trajectory, consistent with the
RG flow itself being unique. We have exemplified this analytically with the
simple BKT model, and shown the relevance of multiple solutions to high energy
physics, where each solution generally corresponds to distinct phenomenology.

Imposing boundary conditions on the CMSSM naturally involves different
renormalisation scales, namely 
the weak scale, $\msusy$, and the gauge coupling unification scale. Thus, the
CMSSM is a good candidate for a theory possessing several different
solutions, for a single set of input parameters sign$(\mu)$, $m_0$, $M_{1/2}$,
$A_0$ and $\tan \beta$. Publicly available computer programs solve for
the RG flow using iteration, which finds at most one
solution consistent with the boundary conditions, for a given point in the CMSSM
parameter space. However, we have 
shown that multiple, physically distinct solutions  
 exist in extended regions of parameter space due to the multiple boundary
 nature of the problem\footnote{This is in contrast to multiple solutions
   found in a more constrained model (mSUGRA)~\cite{Drees:1991ab}, which are
   due to a single non-linear boundary condition.}. 
We showed this by 
relaxing a boundary condition, scanning over the parameter $\mu(\msusy)$, and
(for reasons of computational efficiency) completing the rest of the
calculation by 
iteration. We find
additional, previously unknown and unexplored solutions at different values of
$\mu(\msusy)$, which we 
cryptically referred to in the title as ``the dark
side of the $\mu$''. However, we might well
have found yet more solutions if we had
scanned over other parameters, so in reality all of the parameters
that are not directly fixed by a BC have just as
much of a dark side, waiting to be explored. 

In order to be sure of finding all of the solutions, one could turn the boundary value problem
into an initial value problem and perform the RG flow in one direction only.
One possible method would be to start at the GUT scale and, for a given
sign$(\mu)$, $m_0$, $A_0$,  $M_{1/2}$, and $\tan \beta(\mbox{input})$,
use $\mgut$, $\tan \beta(\mgut)$, 
$g_1(\mgut)=g_2(\mgut)$, $g_3(\mgut)$, $h_t(\mgut)$, $h_b(\mgut)$,
$h_\tau(\mgut)$, $\mu(\mgut)$, $m_3^2(\mgut)$, $v(\mgut)$
and $\msusy$ as scanning parameters.
This 11-dimensional scan, if performed finely enough, would then find all of
the possible solutions that match 
$g_1(\mbox{exp})$, $g_2(\mbox{exp})$, $\msusy$, $m_3^2(\msusy)$,
$g_3(\mbox{exp})$, $v(M_Z)$, $m_b$, $m_t$, $m_\tau$, $M_Z$ and $\tan
\beta(M_Z)$.\footnote{$\tan
\beta(M_Z)$ is, strictly speaking, an output of our
scan. Nevertheless, we follow convention and include it in the above
list of input parameters for the CMSSM.} Note that we found that even while
scanning only in one 
dimension  and 
finding the rest of the parameters by iteration, some of the
low-energy output
parameters are rapidly varying functions of the high-energy inputs, so that finding some of the solutions is
computationally 
intensive. This problem would inevitably be more acute for an 11-dimensional 
scan, 
and we conclude that it would be extremely difficult to know for sure whether
one has 
found all of the solutions, in the absence of a theorem as to their
number.  

% START CHANGES
Here, we have shown examples of phenomenologically viable multiple
solutions where some sparticle masses differ by a few per cent
between the solutions.  However, we have also found other solutions
where some sparticle masses differ by order 100\%; for example,
the 4 solutions in Fig.~\ref{fig:multiPheno-muvsm0} where $m_0\sim3500$ GeV
have this property.  Due to light neutralinos and/or charginos, these
solutions are phenomenologically excluded, but their existence points
to the possibility of phenomenologically viable solutions with large
differences in their spectra.
% END CHANGES

It seems at least plausible, then, that some regions of CMSSM parameter space have been ruled
out erroneously, because of the existence of additional solutions,
yet to be found, with
a phenomenology that is consistent with current bounds. 
For example, we have seen that neutralinos and chargino masses may differ at the tree-level
between the multiple
solutions, and so bounds coming from dark matter searches 
are liable to change, since they are very sensitive to the mass of the
dark matter candidate (in this case, the lightest
neutralino). Similarly, we have seen in
Fig.~\ref{fig:multiPheno-mh0vsm0} that the same parameter point can
lead to multiple predictions for
$m_{h^0}$; in the coming era of precision Higgs physics, we may well
find points for which the solution found by the
standard algorithms is incompatible with the
measured $m_h$, while other solutions are compatible. 

Should we therefore consider all existing exclusions of the CMSSM
parameter space as suspect? Fortunately, there are sectors in which we do not expect the phenomenology
to be greatly changed for different solutions, such that existing
exclusions from experimental searches may be considered robust. 
Important examples are collider searches for
gluinos and squarks decaying into jets and missing energy, which provide
powerful 
constraints upon the CMSSM~\cite{ATLAS-CONF-2012-109,Chatrchyan:2012lia}. 
The gluino/squark production cross-sections relevant for the calculation of
these bounds depend upon the gluino and first
two-family squark masses. 
Since these are fixed at the GUT scale by the boundary condition 
$M_3(\mgut)=M_{1/2}$ and $m_{\tilde q}(\mgut)=m_0$, they only differ 
between the different solutions at the loop level. Perturbativity implies that
these corrections
are small, and so one expects that the production cross-sections will not
differ greatly between the various solutions. Thus, we expect the
previously calculated limits on the CMSSM from squark and gluino production to
approximately hold. One significant {\em caveat}\/ to this is in regions of
parameter 
space where the ratio of the 
  neutralino mass to the gluino or squark mass 
  differs significantly between solutions. This can significantly
  change the kinematics 
  of the decay of the gluino and squark, and therefore can affect the
  acceptance of signal events passing the cuts applied in the
  experimental  analyses.  

For the time being, in the absence of a signal for supersymmetry at
colliders, multiple solutions add a potential loop hole to some of the exclusion
bounds derived in
models such as the CMSSM\@. If evidence for
supersymmetry is found in the future, the phenomenon will present a
new facet: points in the parameter space of candidate SUSY theories
might be incorrectly discarded on the 
grounds of not being able to explain the signal, when other, unknown solutions,
corresponding to the same parameter point, can explain it. We
thus might be led in the wrong direction in searching for the correct
theoretical description of new physics. Conversely, if all the
solutions are known, then we should be able to discriminate between
them given sufficient measurements, given that they are physically
distinct. 

There has been a recent growing industry in multi-dimensional phenomenological
fits of parameter space in the CMSSM to collider and astrophysical data (see
Refs.~\cite{Allanach:2005kz,Buchmueller:2011ab,Balazs:2012qc,Cabrera:2012vu,Fowlie:2012im,Strege:2012bt}
for some examples). These fits use either Bayesian or 
frequentist statistics. The Bayesian 
fits yield posterior probability densities, which are weighted by the
probability masses in marginalised parameters. The addition of multiple
CMSSM solutions to these fits would only have a negligible affect if the
probability mass associated with the non-standard solutions is negligible. 
Performing a Bayesian fit with the additional 11 dimensions of our scan
would be an interesting exercise, to see what the effect of the multiple
solutions is. Such fits, since they scan in $m_3^2$ and $\mu$ (among other
parameters) rather than
$\tan \beta$ (and fixing $\mu$ by $M_Z(\mbox{exp})$), will implicitly and
automatically take into account a Jacobian factor that was calculated for this
purpose~\cite{Allanach:2007qk,Cabrera:2008tj}. 
Frequentist fits must scan in the additional solution
space and may be even more vulnerable to changes if they have a better
best-fit point, since parameter estimation always relies on $\Delta \chi^2$,
the difference in $\chi^2$ between the current parameter space point and the
best-fit point. It remains to be seen how much these effects will change the
fits in each case, but at the moment they remain incomplete without the
inclusion of the multiple solutions exhibited here.

Finally, we emphasise that the CMSSM is only one example of a model where the
SUSY breaking conditions are mostly set at a scale much higher than the weak
scale; we expect other examples to feature multiple
solutions as well.

%%%%%%%%%%%%%%%%%%%%%%%%%%%%%%%%%%%%%%%%%%%%%%%%%%%%%%%%%%%%%

%%%%%%%%%%%%%%%%%%%%%%%%%%%%%%%%%%%%%%%%%%%%%%%%%%%%%%%%%
\section*{Acknowledgements}
%%%%%%%%%%%%%%%%%%%%%%%%%%%%%%%%%%%%%%%%%%%%%%%%%%%%%%%%
This work has been partially supported by STFC.
DG is funded by a Herchel Smith fellowship.
We would like to thank other members of the Cambridge SUSY
Working Group and C.~Mouhot for discussions, and S.~Thorgerson ({\em requiescat in pace}\/) for
inspiration.  

\appendix

\section{Standard Model input parameters \label{sec:inputs}}
As input parameters, we have~\cite{PDG} 
$G_F=1.16637$ GeV$^{-2}$,
pole masses $m_t=173.5$
GeV, $M_Z(\mbox{exp})=91.1887$ GeV and $m_\tau=1.77669$ GeV, modified minimal
subtraction scheme Standard Model values of: $m_b(m_b)=4.18$ GeV,
$\alpha (M_Z)^{-1}=127.916$, $\alpha_s(M_Z)=0.1187$. 

\bibliographystyle{JHEP-2}
\bibliography{RGE}

\providecommand{\href}[2]{#2}\begingroup\raggedright\begin{thebibliography}{10}

\bibitem{Lindelhof}
{Lindel\"{o}f, E.}, {\it {Sur l'application de la m\'{e}thode des
  approximations successives aux \'{e}quations diff\'{e}rentielles ordinaires
  du premier ordre}},  {\em Comptes Rendus} {\bf 114} (1894) 454.

\bibitem{Liu:2013ula}
M.~Liu and P.~Nath, {\it {Higgs Boson Mass, Proton Decay, Naturalness and
  Constraints of LHC and Planck Data}},
  \href{http://arXiv.org/abs/1303.7472}{{\tt 1303.7472}}.
%%CITATION = ARXIV:1303.7472;%%

\bibitem{Berezinsky:1970fr}
V.~Berezinsky, {\it {Destruction of long range order in one-dimensional and
  two-dimensional systems having a continuous symmetry group. 1. Classical
  systems}},  {\em Sov.Phys.JETP} {\bf 32} (1971) 493--500.
%%CITATION = SPHJA,32,493;%%

\bibitem{Kosterlitz:1973xp}
J.~Kosterlitz and D.~Thouless, {\it {Ordering, metastability and phase
  transitions in two-dimensional systems}},  {\em J.Phys.} {\bf C6} (1973)
  1181--1203.
%%CITATION = INSPIRE-81176;%%

\bibitem{Polyakov:1976fu}
A.~M. Polyakov, {\it {Quark Confinement and Topology of Gauge Groups}},  {\em
  Nucl.Phys.} {\bf B120} (1977) 429--458.
%%CITATION = NUPHA,B120,429;%%

\bibitem{Svetitsky:1982gs}
B.~Svetitsky and L.~G. Yaffe, {\it {Critical Behavior at Finite Temperature
  Confinement Transitions}},  {\em Nucl.Phys.} {\bf B210} (1982) 423.
%%CITATION = NUPHA,B210,423;%%

\bibitem{Agasian:1997wv}
N.~O. Agasian and K.~Zarembo, {\it {Phase structure and nonperturbative states
  in three-dimensional adjoint Higgs model}},  {\em Phys.Rev.} {\bf D57} (1998)
  2475--2485 [\href{http://arXiv.org/abs/hep-th/9708030}{{\tt
  hep-th/9708030}}].
%%CITATION = HEP-TH/9708030;%%

\bibitem{Gripaios:2002xb}
B.~Gripaios, {\it {Variational analysis of deconfinement in compact U(1) gauge
  theory}},  {\em Phys.Rev.} {\bf D67} (2003) 025023
  [\href{http://arXiv.org/abs/hep-th/0211104}{{\tt hep-th/0211104}}].
%%CITATION = HEP-TH/0211104;%%

\bibitem{Fayet:1}
P.~Fayet, {\it {Supersymmetry and Weak, Electromagnetic and Strong
  Interactions}},  {\em Phys.Lett.} {\bf B64} (1976) 159.

\bibitem{Fayet:2}
P.~Fayet, {\it {Spontaneously Broken Supersymmetric Theories of Weak,
  Electromagnetic and Strong Interactions}},  {\em Phys.Lett.} {\bf B69} (1977)
  489.

\bibitem{Fayet:3}
G.~R. Farrar and P.~Fayet, {\it {Phenomenology of the Production, Decay and
  Detection of New Hadronic States Associated with Supersymmetry}},  {\em
  Phys.Lett.} {\bf B76} (1978) 575.

\bibitem{Fayet:4}
P.~Fayet, {\it {Relations Between the Masses of the Superpartners of Leptons
  and Quarks, the Goldstino Couplings and the Neutral Currents}},  {\em
  Phys.Lett.} {\bf B84} (1979) 416.

\bibitem{Georgi:1}
S.~Dimopoulos and H.~Georgi, {\it {Softly Broken Supersymmetry and $SU(5)$}},
  {\em Nucl.Phys.} {\bf B193} (1981) 150.

\bibitem{Cremmer:1978iv}
E.~Cremmer, B.~Julia, J.~Scherk, P.~van Nieuwenhuizen, S.~Ferrara {\em
  et.~al.}, {\it {SuperHiggs Effect in Supergravity with General Scalar
  Interactions}},  {\em Phys.Lett.} {\bf B79} (1978) 231.
%%CITATION = PHLTA,B79,231;%%

\bibitem{Cremmer:1978hn}
E.~Cremmer, B.~Julia, J.~Scherk, S.~Ferrara, L.~Girardello {\em et.~al.}, {\it
  {Spontaneous Symmetry Breaking and Higgs Effect in Supergravity Without
  Cosmological Constant}},  {\em Nucl.Phys.} {\bf B147} (1979) 105.
%%CITATION = NUPHA,B147,105;%%

\bibitem{Barbieri:1982eh}
R.~Barbieri, S.~Ferrara and C.~A. Savoy, {\it {Gauge Models with Spontaneously
  Broken Local Supersymmetry}},  {\em Phys.Lett.} {\bf B119} (1982) 343.
%%CITATION = PHLTA,B119,343;%%

\bibitem{Drees:1991ab}
M.~Drees and M.~M. Nojiri, {\it {Radiative symmetry breaking in minimal $N=1$
  supergravity with large Yukawa couplings}},  {\em Nucl.Phys.} {\bf B369}
  (1992) 54--98.
%%CITATION = NUPHA,B369,54;%%

\bibitem{Yamada}
Y.~Yamada, {\it {From minimal to realistic supersymmetric SU(5) grand
  unification}},  {\em Z.\ Phys.\ C} {\bf 60} (1993) 83--94.

\bibitem{Altarelli:2000fu}
G.~Altarelli, F.~Feruglio and I.~Masina, {\it {From minimal to realistic
  supersymmetric SU(5) grand unification}},  {\em JHEP} {\bf 0011} (2000) 040
  [\href{http://arXiv.org/abs/hep-ph/0007254}{{\tt hep-ph/0007254}}].
%%CITATION = HEP-PH/0007254;%%

\bibitem{AbdusSalam:2011fc}
S.~AbdusSalam, B.~Allanach, H.~Dreiner, J.~Ellis, U.~Ellwanger {\em et.~al.},
  {\it {Benchmark Models, Planes, Lines and Points for Future SUSY Searches at
  the LHC}},  {\em Eur.Phys.J.} {\bf C71} (2011) 1835
  [\href{http://arXiv.org/abs/1109.3859}{{\tt 1109.3859}}].
%%CITATION = ARXIV:1109.3859;%%

\bibitem{Martin:1993zk}
S.~P. Martin and M.~T. Vaughn, {\it {Two loop renormalization group equations
  for soft supersymmetry breaking couplings}},  {\em Phys.Rev.} {\bf D50}
  (1994) 2282 [\href{http://arXiv.org/abs/hep-ph/9311340}{{\tt
  hep-ph/9311340}}].
%%CITATION = HEP-PH/9311340;%%

\bibitem{Capper:1979ns}
D.~Capper, D.~Jones and P.~van Nieuwenhuizen, {\it {Regularization by
  Dimensional Reduction of Supersymmetric and Nonsupersymmetric Gauge
  Theories}},  {\em Nucl.Phys.} {\bf B167} (1980) 479.
%%CITATION = NUPHA,B167,479;%%

\bibitem{Allanach:2001kg}
B.~Allanach, {\it {SOFTSUSY: a program for calculating supersymmetric
  spectra}},  {\em Comput.Phys.Commun.} {\bf 143} (2002) 305--331
  [\href{http://arXiv.org/abs/hep-ph/0104145}{{\tt hep-ph/0104145}}].
%%CITATION = HEP-PH/0104145;%%

\bibitem{Baer:1993ae}
H.~Baer, F.~E. Paige, S.~D. Protopopescu and X.~Tata, {\it {Simulating
  Supersymmetry with ISAJET 7.0 / ISASUSY 1.0}},
  \href{http://arXiv.org/abs/hep-ph/9305342}{{\tt hep-ph/9305342}}.
%%CITATION = HEP-PH/9305342;%%

\bibitem{Porod:2003um}
W.~Porod, {\it {SPheno, a program for calculating supersymmetric spectra, SUSY
  particle decays and SUSY particle production at e+ e- colliders}},  {\em
  Comput.Phys.Commun.} {\bf 153} (2003) 275--315
  [\href{http://arXiv.org/abs/hep-ph/0301101}{{\tt hep-ph/0301101}}].
%%CITATION = HEP-PH/0301101;%%

\bibitem{Chowdhury:2011zr}
D.~Chowdhury, R.~Garani and S.~K. Vempati, {\it {SUSEFLAV: Program for
  supersymmetric mass spectra with seesaw mechanism and rare lepton flavor
  violating decays}},  {\em Comput.Phys.Commun.} {\bf 184} (2013) 899--918
  [\href{http://arXiv.org/abs/1109.3551}{{\tt 1109.3551}}].
%%CITATION = ARXIV:1109.3551;%%

\bibitem{Djouadi:2002ze}
A.~Djouadi, J.-L. Kneur and G.~Moultaka, {\it {SuSpect: A Fortran code for the
  supersymmetric and Higgs particle spectrum in the MSSM}},  {\em
  Comput.Phys.Commun.} {\bf 176} (2007) 426--455
  [\href{http://arXiv.org/abs/hep-ph/0211331}{{\tt hep-ph/0211331}}].
%%CITATION = HEP-PH/0211331;%%

\bibitem{Pierce:1996zz}
D.~M. Pierce, J.~A. Bagger, K.~T. Matchev and R.-j. Zhang, {\it {Precision
  corrections in the minimal supersymmetric standard model}},  {\em Nucl.Phys.}
  {\bf B491} (1997) 3--67 [\href{http://arXiv.org/abs/hep-ph/9606211}{{\tt
  hep-ph/9606211}}].
%%CITATION = HEP-PH/9606211;%%

\bibitem{ATLAS:2012oga}
{\bf ATLAS Collaboration} Collaboration, G.~Aad {\em et.~al.}, {\it {A particle
  consistent with the Higgs Boson observed with the ATLAS Detector at the Large
  Hadron Collider}},  {\em Science} {\bf 338} (2012) 1576--1582.
%%CITATION = SCIEA,338,1576;%%

\bibitem{Chatrchyan:2012ufa}
{\bf CMS Collaboration} Collaboration, S.~Chatrchyan {\em et.~al.}, {\it
  {Observation of a new boson at a mass of 125 GeV with the CMS experiment at
  the LHC}},  {\em Phys.Lett.} {\bf B716} (2012) 30--61
  [\href{http://arXiv.org/abs/1207.7235}{{\tt 1207.7235}}].
%%CITATION = ARXIV:1207.7235;%%

\bibitem{Allanach:2000ii}
B.~Allanach, J.~Hetherington, M.~A. Parker and B.~Webber, {\it {Naturalness
  reach of the large hadron collider in minimal supergravity}},  {\em JHEP}
  {\bf 0008} (2000) 017 [\href{http://arXiv.org/abs/hep-ph/0005186}{{\tt
  hep-ph/0005186}}].
%%CITATION = HEP-PH/0005186;%%

\bibitem{Allanach:2012qd}
B.~Allanach and M.~Parker, {\it {Uncertainty in Electroweak Symmetry Breaking
  in the Minimal Supersymmetric Standard Model and its Impact on Searches For
  Supersymmetric Particles}},  {\em JHEP} {\bf 1302} (2013) 064
  [\href{http://arXiv.org/abs/1211.3231}{{\tt 1211.3231}}].
%%CITATION = ARXIV:1211.3231;%%

\bibitem{ATLAS-CONF-2012-109}
{\it Search for squarks and gluinos with the atlas detector using final states
  with jets and missing transverse momentum and 5.8 fb$^{-1}$ of $\sqrt{s}$=8
  tev proton-proton collision data},  Tech. Rep. ATLAS-CONF-2012-109, CERN,
  Geneva, Aug, 2012.

\bibitem{Chatrchyan:2012lia}
{\bf CMS Collaboration} Collaboration, S.~Chatrchyan {\em et.~al.}, {\it
  {Search for new physics in the multijet and missing transverse momentum final
  state in proton-proton collisions at $\sqrt{s} = 7$ TeV}},  {\em
  Phys.Rev.Lett.} {\bf 109} (2012) 171803
  [\href{http://arXiv.org/abs/1207.1898}{{\tt 1207.1898}}].
%%CITATION = ARXIV:1207.1898;%%

\bibitem{Allanach:2005kz}
B.~Allanach and C.~Lester, {\it {Multi-dimensional mSUGRA likelihood maps}},
  {\em Phys.Rev.} {\bf D73} (2006) 015013
  [\href{http://arXiv.org/abs/hep-ph/0507283}{{\tt hep-ph/0507283}}].
%%CITATION = HEP-PH/0507283;%%

\bibitem{Buchmueller:2011ab}
O.~Buchmueller, R.~Cavanaugh, A.~De~Roeck, M.~Dolan, J.~Ellis {\em et.~al.},
  {\it {Higgs and Supersymmetry}},  {\em Eur.Phys.J.} {\bf C72} (2012) 2020
  [\href{http://arXiv.org/abs/1112.3564}{{\tt 1112.3564}}].
%%CITATION = ARXIV:1112.3564;%%

\bibitem{Balazs:2012qc}
C.~Balazs, A.~Buckley, D.~Carter, B.~Farmer and M.~White, {\it {Should we still
  believe in constrained supersymmetry?}},
  \href{http://arXiv.org/abs/1205.1568}{{\tt 1205.1568}}.
%%CITATION = ARXIV:1205.1568;%%

\bibitem{Cabrera:2012vu}
M.~E. Cabrera, J.~A. Casas and R.~R. de~Austri, {\it {The health of SUSY after
  the Higgs discovery and the XENON100 data}},
  \href{http://arXiv.org/abs/1212.4821}{{\tt 1212.4821}}.
%%CITATION = ARXIV:1212.4821;%%

\bibitem{Fowlie:2012im}
A.~Fowlie, M.~Kazana, K.~Kowalska, S.~Munir, L.~Roszkowski {\em et.~al.}, {\it
  {The CMSSM Favoring New Territories: The Impact of New LHC Limits and a 125
  GeV Higgs}},  {\em Phys.Rev.} {\bf D86} (2012) 075010
  [\href{http://arXiv.org/abs/1206.0264}{{\tt 1206.0264}}].
%%CITATION = ARXIV:1206.0264;%%

\bibitem{Strege:2012bt}
C.~Strege, G.~Bertone, F.~Feroz, M.~Fornasa, R.~R. de~Austri {\em et.~al.},
  {\it {Global Fits of the cMSSM and NUHM including the LHC Higgs discovery and
  new XENON100 constraints}},  {\em JCAP} {\bf 1304} (2013) 013
  [\href{http://arXiv.org/abs/1212.2636}{{\tt 1212.2636}}].
%%CITATION = ARXIV:1212.2636;%%

\bibitem{Allanach:2007qk}
B.~C. Allanach, K.~Cranmer, C.~G. Lester and A.~M. Weber, {\it {Natural priors,
  CMSSM fits and LHC weather forecasts}},  {\em JHEP} {\bf 0708} (2007) 023
  [\href{http://arXiv.org/abs/0705.0487}{{\tt 0705.0487}}].
%%CITATION = ARXIV:0705.0487;%%

\bibitem{Cabrera:2008tj}
M.~Cabrera, J.~Casas and R.~Ruiz~de Austri, {\it {Bayesian approach and
  Naturalness in MSSM analyses for the LHC}},  {\em JHEP} {\bf 0903} (2009) 075
  [\href{http://arXiv.org/abs/0812.0536}{{\tt 0812.0536}}].
%%CITATION = ARXIV:0812.0536;%%

\bibitem{PDG}
{\bf Particle Data Group} Collaboration, J.~Beringer {\em et.~al.}, {\it
  {Review of Particle Physics (RPP)}},  {\em Phys.Rev.} {\bf D86} (2012)
  010001.
%%CITATION = PHRVA,D86,010001;%%

\end{thebibliography}\endgroup

\end{document}